\documentclass[12pt,preprint]{aastex}

\begin{document}
\newcommand{\mjup}{M$_{\rm Jup}$}
\newcommand{\mearth}{M$_{\oplus}$}

\title{Forming Close-in Earth-like Planets via a Collision-Merger
Mechanism in  Late-stage Planet Formation}

\author{Jianghui Ji\altaffilmark{1,2}, Sheng Jin\altaffilmark{1,3}, C. G. Tinney\altaffilmark{2}}%

\altaffiltext{1}{Purple Mountain Observatory, Chinese Academy of
Sciences, Nanjing 210008, China; jijh@pmo.ac.cn}

\altaffiltext{2}{Department of Astrophysics, School of Physics,
University of New South Wales, NSW 2052, Australia}

\altaffiltext{3}{Graduate School of Chinese Academy of Sciences,
Beijing 100049, China; qingxiaojin@gmail.com}

\begin{abstract}

The large number of exoplanets found to orbit their host stars in
very close orbits have significantly advanced our understanding of
the planetary formation process. It is now widely accepted that such
short-period planets cannot have formed {\em in situ}, but rather
must have migrated to their current orbits from a formation location
much farther from their host star. In the late stages of planetary
formation, once the gas in the proto-planetary disk has dissipated
and migration has halted, gas-giants orbiting in the inner disk
regions will excite planetesimals and planetary embryos, resulting
in an increased rate of orbital crossings and large impacts. We
present the results of dynamical simulations for planetesimal
evolution in this later stage of planet formation. We find that a
mechanism is revealed by which the collision-merger of planetary
embryos can kick terrestrial planets directly into orbits extremely
close to their parent stars.

\end{abstract}


\keywords{celestial mechanics -- methods:numerical -- planets and
satellites:formation -- stars:individual (OGLE-06-109L, 47 Ursae
Majoris)}

\section{INTRODUCTION}
To date over 490 extrasolar planets have been discovered, revealing
a wide diversity of planetary systems (http://exoplanet.eu). One of
more unusual phenomena so revealed has been the population of ``Hot
Jupiters'' -- gas-giants found in very small orbits (periods $<$ 8d)
about their parent stars -- of which the prototype was the very
first gas-giant exoplanet discovered, 51\,Peg \citep{may95}. It is
believed that such short-period gas-giants cannot have formed this
close to their parent stars, and so must have migrated in, or been
scattered in, from a more distant formation region
\citep{lin96,weid96,Ida04,cham09}. The measurement precisions that
make the detection of such short-period exoplanets possible have
over recent years continually improved for both Doppler (e.g. Gl 876
d \citep{rive05}, Gl 581 c \citep{udry07}, 61 Vir b \citep{vogt10})
and transit (e.g. Kepler-4b \citep{boru10}) detection. What then are
the possible formation mechanisms that can produce such close-in
terrestrial and super-terrestrial planets?

Several models have been proposed for the formation of close-in
terrestrial planets. \citet{raym06} have shown that super-Earths could
form interior to a migrating Jovian planet. As they migrate inward,
such gas-giants can shepherd planetary embryos interior to their orbits,
which can then further collide and merge to generate Earth-like planets
\citep{zhou05}.
It has also been suggested that orbital migration and
planet-planet scattering could potentially produce
short-period super-Earths \citep{brun05, terq07, raym08}.
Whatever the mechanism for their formation, it is likely that
such planets are common around at least low-mass stars \citep{kenn08}.

In all these scenarios, the formation of short-period Earth-like
planets is associated with the migration of gas-giant planets.
According to the core accretion paradigm for planetary formation,
the isolation cores in the terrestrial planet formation region, and
the solid cores of gas-giants, are both formed within $\sim 1$\,Myr
from kilometer-sized planetesimals \citep{Saf69,wet80}. Subsequently
massive solid cores accrete disk gas to form giant planets
\citep{Kok02,Ida04} at $\sim 3-6$\,Myr, before the disk disperses
\citep{hais01}. In the late stage of planet formation, when giant
planets have ceased migration after the gas disk clears, the disk of
countless planetesimals and planetary embryos will become turbulent
due to stirring by gas-giants over hundreds of million years (or
potentially even longer). In the meantime, it is expected that
orbital crossings and giant impacts will frequently occur, which
could lead to the formation of terrestrial planets
\citep{cham01,raym04,zha09} and short-period Earth-like planets.

In this Letter, we present a potential new formation mechanism for
short-period Earth-like planets in the late stage of planet
formation through a collision-merger scenario. In this mechanism, a
planetary embryo is directly kicked to a close-in orbit after a
collision with another embryo, and then the larger merged body is
seized by the central star as a hot Earth-like planet.

\section{SIMULATION SETUP}

Extrasolar planetary systems that harbor pairs of
Jupiter-to-Saturn-mass companions are of particular interest to
researchers \citep{Goz02,Ji05,zha10}, e.g., OGLE-06-109L\,bc
\citep{gaud08}, 47\,Uma\,bc \citep{butl96,fish02}, Gl\,876\,bc
\citep{marc01}. It is interesting to consider whether it is likely
that such systems might host additional hot terrestrial planets (as,
for example, the Gl\,876 system does in the form of Gl\,876\,d --
\citet{rive05}), and further how such planets might form and evolve.
We have therefore carried out simulations that explore such a system
architecture.

In total, 30 runs were performed using a hybrid symplectic algorithm
in the MERCURY package \citep{cham99} for following two such
systems. The initial conditions of the two systems simulated were:

\begin{list}{}{\itemsep=-\parsep}
\item[\bf Simulation 1] - two giant planets are simulated with
initial orbital parameters ($M_{P}$, $a$, $e_{p}$) $=$ (0.71\,\mjup,
2.3\,AU, 0.001) and (0.27\,\mjup, 4.6\,AU, 0.11), to emulate the
OGLE-2006-BLG-109L system \citep{gaud08}. 500 planetary embryos and
planetesimals \footnote{Herein the masses of embryos range from
several lunar-mass to Mar-mass, and those of smaller "planetesimals"
have approximately a lunar mass, rather than a planetesimal mass.}
with total mass 10\,\mearth\ were distributed between 0.3\,AU $< a
<$ 5.2\,AU and with $e<0.02$. Each of the 26 runs carried out over
400\,Myr.

\item[\bf Simulation 2] - two giant planets are simulated with
initial orbital parameters  ($M_{P}$, $a$, $e_{p}$) $=$ (2.9\,\mjup,
2.08\,AU, 0.05) and (1.1\,\mjup, 3.97\,AU, 0.001), to emulate the
47\,Uma system \citep{fish02}. 648 planetary objects with total mass
of 5.14\,\mearth were distributed in the region 0.3\,AU $ < a < $
1.6\,AU with $e < 0.02$. Each of the four runs evolved over
100\,Myr.
\end{list}

The other initial orbital elements of each planetary embryo (or
planetesimal) are randomly generated -- the arguments of periastron,
longitudes of the ascending node, and mean anomalies range from
$0^{\circ}$ to $360^{\circ}$, and inclinations are from $0^{\circ}$
to $1^{\circ}$. In addition, the hybrid integrator parameters are
adopted as a stepsize of $3$ days ($\sim$ a twentieth of a period
for the innermost possible body at $0.3$\,AU), and a Bulirsch-Stoer
tolerance of $10^{-12}$. At the end of integration, the changes of
energy and angular momenta are $10^{-3}$ and $10^{-11}$,
respectively. In these runs, the gravitational interactions of all
bodies are taken into account. Two bodies are assumed to collide
whenever the distance between them is less than the sum of their
physical radii \citep{cham99}. If two objects collide, they are
merged into a single body, without fragmentation, after the
collision.

\section{RESULTS}
\subsection{Simulation results}

\begin{figure*}
\includegraphics[width=17.5cm]{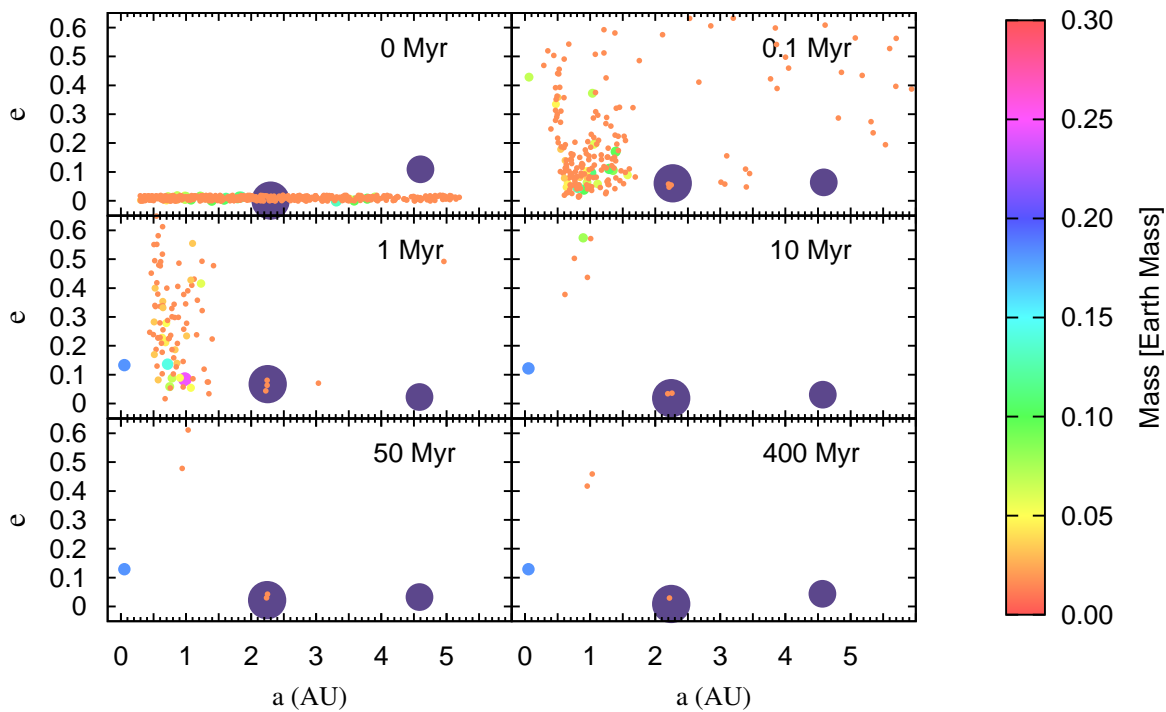}\\
\caption{\label{fig:wide} A snapshot of planet formation in the late
stage for {\bf Simulation 1}. The panels show the orbital eccentricity versus
semi-major axis for each surviving body at simulation times of
0, 0.1, 1.0, 10, 50 \& 400\,Myr. The radii and the color of
the embryos and planetesimals are related to their mass, with radius
proportional to $m^{1/3}$. The two giants are, respectively, at
2.3 and 4.6\,AU. A close-in terrestrial planet forms at $\sim$ 1\,Myr and it
remains stable over secular evolution.}
\end{figure*}

In our simulations, we find that the collision-merger mechanism
produces close-in terrestrial planets in 20\% of the runs carried
out (5 of 26 {\bf Simulation 1} runs, and 1 of 4 {\bf Simulation 2}
runs). The simulations exhibit a classical planetary accretion
scenario in their late stage formation \citep{cham01,raym04}. Figure
\ref{fig:wide} shows snapshots at various evolution times for a
representative run of {\bf Simulation 1}. Initially, the embryos and
planetesimals reside in a cold disk, which is quickly stirred by the
two gas-giants and excited to highly eccentric orbits within
0.1\,Myr. We  also see that three small bodies are involved in a 1:1
resonance with the inner giant by that time. By the end of 1\,Myr,
most of the initial objects have been removed by ejection or
collision due to frequent orbital crossings in this chaotic stage.
In addition, we see that a close-in planet has formed at $\sim$
1\,Myr which subsequently remains very stable. At the conclusion of
the run (400\,Myr), three planetesimals survive, of which one has
been seized as a Trojan body by the inner giant, and the other two
move at $\sim 1$\,AU in eccentric orbits.

\begin{figure}
 \figurenum{2}
   \plotone{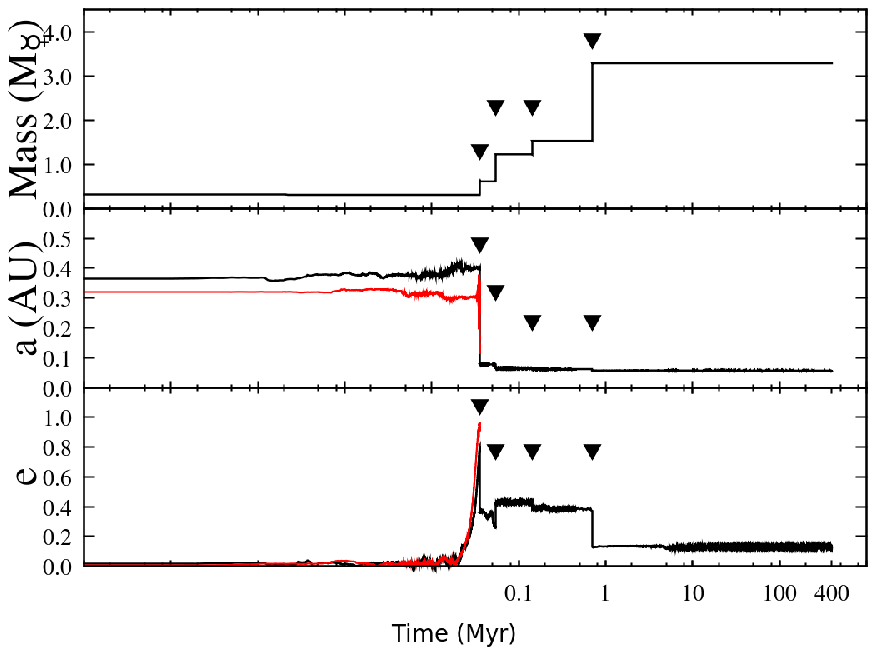}
  \caption{Mass, semi-major axis and eccentricity evolution of the
short-period terrestrial planet the emerges from the {\bf Simulation 1} run shown in
Fig. \ref{fig:wide}. The black and red lines in the lower two panels
show the semi-major axis and eccentricity evolution of the two bodies that collide to
form a merged planetary embryo, which is kicked
 from 0.4\,AU to 0.077\,AU at 0.0356\,Myr. Subsequently that merged embryo (shown as a single
 black line after 0.0356\,Myr) is subject to further collision-mergers, with
 the epoch of each collision shown by the {\em solid triangles}. The resultant
 mass evolution of this body is shown in the upper panel.} \label{fig2}
\end{figure}

Figure \ref{fig2} shows the time evolution of the mass, semi-major
axis, and eccentricity of the short-period terrestrial planet formed
in the {\bf Simulation 1} run shown in  Fig. \ref{fig:wide}. At
0.0356\,Myr, two bodies that may be excited by secular resonance of
gas-giants, collide at very high eccentricities ($e$ = 0.91 and
0.80, shown by the red and black lines in Fig. \ref{fig2},
respectively) and are then assumed to merge into a single planetary
embryo. That merged body (the remaining black line in Fig.
\ref{fig2}) is captured by the parent star as a short-period planet,
and its orbit dramatically shrinks from $\sim 0.4$\,AU at the time
of the collision, down to 0.077\,AU. Subsequently, three additional
collisions take place over the further late-stage evolution of that
merged object. Fig. \ref{fig2} shows that the embryo moves slightly
inward at each collision, and that its mass also increases.
Moreover, we note that it finally becomes a 3.3\,Mercury-mass planet
with a close-in orbit about 0.056\,AU, and its eccentricity drops
down to $e$=0.13 after the last collision. The orbit may then, of
course, be further circularized by tidal interaction with the star
over even longer timescales.

\begin{figure}
 \figurenum{3}
   \plotone{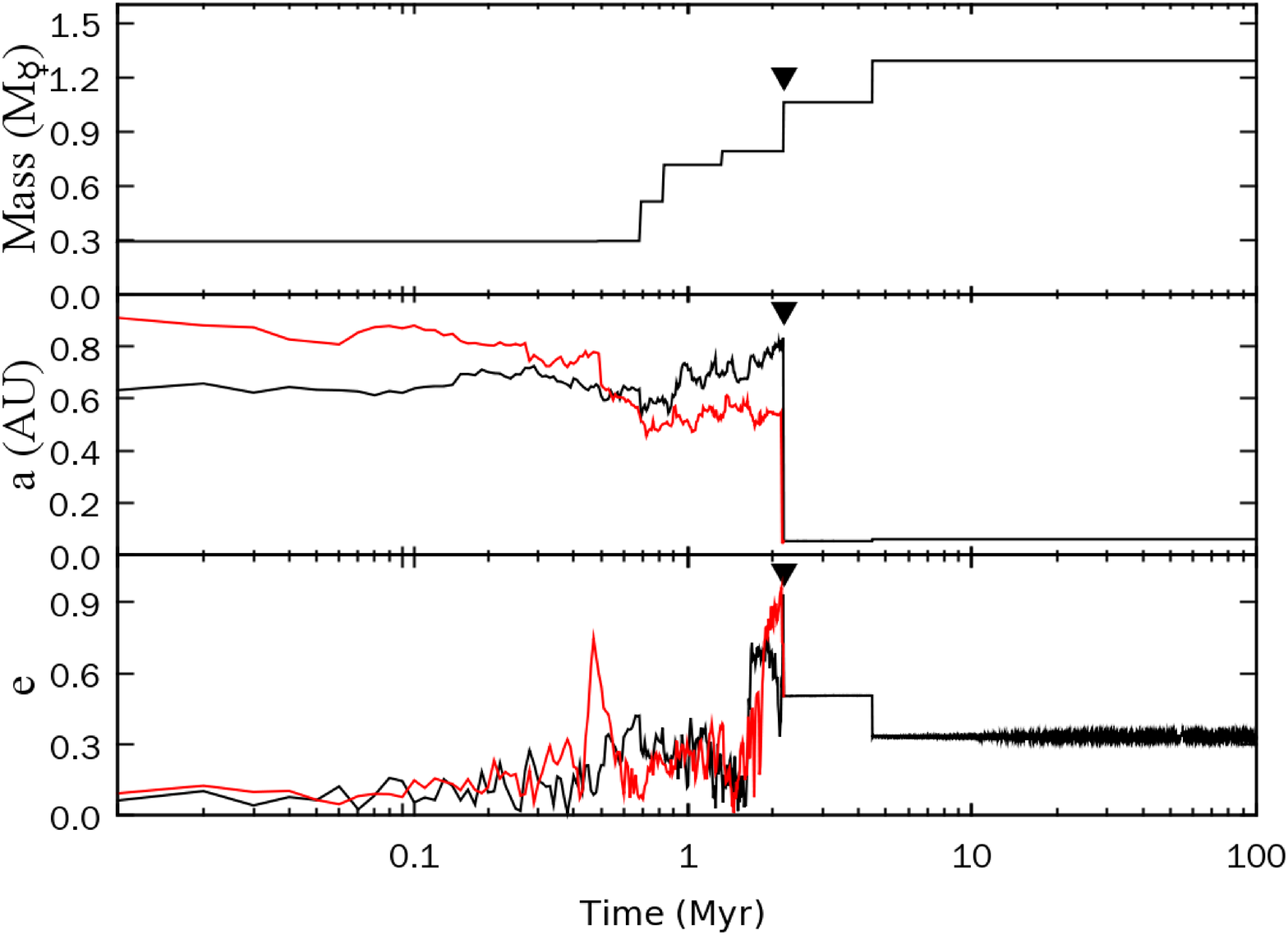}
  \caption{Mass, semi-major axis and eccentricity evolution of a
short-period terrestrial planet that emerges from a run of {\bf
Simulation 2} -- layout is the same as for Fig. \ref{fig2}. In this
case the first collision-merger event occurs at 2.2\,Myr, and the
embryo is thrown from its location of $\sim0.8$\,AU at the time of
the collision to 0.06\,AU.} \label{fig3}
\end{figure}

Figure \ref{fig3} shows the formation and evolution of a similar
terrestrial planet that emerges in one of the runs for {\bf
Simulation 2}. At 2.2\,Myr, the semi-major axis of a $\sim
0.9$\,Mercury-mass embryo drops down from $\sim0.8$\,AU to 0.06\,AU
as a result of a collision with a highly-eccentric planetesimal
excited more than a million years earlier. The merged body has an
eccentricity that drops from 0.90 to 0.50 immediately after the
impact. The enlarged, merged body subsequently undergoes additional
collisions, and its eccentricity further evolves to $e$=0.33 (after
its last collision) with a final mass of 1.3\,Mercury-mass. Here,
the collision-merger scenario may provide some clues of the origins
of the moderate eccentricities seen in super-Earths detected to date
(e.g., HD 181433 b \citep{Bouc09}).

The major difference in the evolution of these two examples is that
the short-period planet that evolved in {\bf Simulation 1} was moved
into an inner orbit at a very early stage, and subsequently accreted
a majority of the mass available in nearby orbits; while the {\bf
Simulation 2} planet had almost completed accretion into a
Mercury-mass embryo before it moved closer to the star. In all
simulations, we notice that terrestrial planets and bodies formed at
short periods via collision-merger events come into being within
10-30 Myr, which agrees with the estimated timescale of terrestrial
core formation \citep{yin02}, as derived from the chronometry of
meteorites and numerical simulations of terrestrial planet formation
\citep{cham01,raym04,zha09}. In addition, we find that all survivors
remain stable in their final configurations.

These results indicate that a collision-merger mechanism could
indeed produce short-period, terrestrial planets in two systems that
host two gas-giants. Similarly, we also find the above outcomes in
other 4 runs. However, a natural question then arises -- do the
bodies that take part in these collisions really merge? Or will they
become fragmented?

\subsection{Merger versus fragmentation}

In the accretion model of MERCURY, a collision-merger scenario
occurs whenever the distance between two bodies is less than the sum
of their physical radii \citep{cham99}, and the package models the
two bodies merging inelastically to form a single new body that
conserves mass and total momentum.  The collisions in the runs,
therefore, are considered to be perfect gravitational aggregations,
which assumes that enough energy is dissipated in the collision for
the two bodies to remain gravitationally bound. However, actual
collisions could have a result that ranges anywhere from this result
(complete merger), through partial fragmentation, to the complete
shattering and disintegration of both impactors. \citep{wet93}.

Whether these bodies either fragment or cohere in a collision will
obviously depend on the -- currently poorly understood --  physical
properties of the colliding bodies \citep{wet80}. What can be said
is that the outcome will be extremely complicated. To assess the
likely state of the merger vs fragmentation for two bodies in a
collision, we can, though, make order-of-magnitude estimates.

Consider two bodies of the same mass $m$, with relative velocity at
infinity $\sigma$, and the sum of the physical radii $R_{s}$. The
collision velocity for a head-on collision between them is
\citep{Saf69,wet80,armi07},
\begin{equation}
v_{c}=(\sigma^{2}+v^{2}_{esc})^{1/2}
\end{equation}
where $v_{esc}$=$\sqrt{4Gm/R_{s}}$ is the escape velocity at the
point of collision, a parameter used to evaluate whether they will
physically collide. Take the coefficient of restitution  as
$\epsilon$, then accretion will result if $\epsilon v_{c}<v_{esc}$,
even if the initial impact results in fragmentation into two bodies.
Conversely, the bodies will be unbound if $\epsilon v_{c}>v_{esc}$.
Thus, the threshold value of the coefficient of restitution for
these outcomes is \citep{armi07},
\begin{equation}
\epsilon=\bigg(1+\frac{\sigma^{2}}{v^{2}_{esc}}\bigg)^{-1/2}
\end{equation}
This shows that if $\sigma\ll v_{esc}$, merger and growth is likely
unless collision is totally elastic; whereas $\sigma\gg v_{esc}$
leads to fragmentation.

For the {\bf Simulation 1} run shown in Fig. \ref{fig2} , we have
used the above equations to assess the outcome of the first
collision as it happens between two identical Mercury-like embryos,
which allows us to  make a rough evaluation of the likely outcome by
calculating the instantaneous velocities of the impactors at the
epoch just before the collision. Now we notice that at the first
collision the body was impacted onto a close-in orbit. The $v_{esc}$
of the two impactors are nearly the same -- 3.12\,km\,s$^{-1}$
(assuming equal bulk density); the velocities of the impactors at
the collision epoch near the pericenter are estimated to be
43.21\,km\,s$^{-1}$ and 35.50\,km\,s$^{-1}$, respectively, thus we
have an approximate relative velocity projected to the relative
position of two colliding bodies of 13.49\,km\,s$^{-1}$. In this
case, a merger requires $\epsilon\leq0.23$, which is close to the
accretion condition ($\epsilon\leq0.34$)  in realistic accretion
model for head-on collision \citep{Kok10}. On the basis of above
analysis, a merger seem to be possible for two eccentric objects
when the collision occurs in the nearby region of central star,
subsequently the merged body is seized by the star at close-in
orbit. In the collision-merger process, moderate energy should be
released, and they could be converted into the internal heat of the
merger in the collision between embryos, e.g., simulations of a
supposed Moon-forming impact show that the collision can deliver
prodigious energy to the Earth, which could lead the proto-Earth to
a mixed solid-melt state \citep{canu08}.\footnote{At the very time
before/after the collision, the fractional energy change due to
integrator was about 9 part in $10^4$. Additional energy loss may
arise from the ejection of other embryos or transfer to the envelope
and core of giant planets \citep{Li10}.}

We also obtain similar results for the {\bf Simulation 2} run shown
in Fig. \ref{fig3}.  In addition, \citet{lein02} showed that a large
mass ratio between two impactors will tend to lead to merger and
aggregation -- the accretion probability is $\sim 60$\% (averaged
over all impact parameters) for average mass ratio of 1:5. This
suggests that the first collision seen in this run, where the mass
ratio of 1:3.43, is likely to result in a merger.

\section{DISCUSSION and CONCLUSION}

We have uncovered a new mechanism for producing short-period
terrestrial planets via collisions-mergers in the late stages of
planetary formation. In this mechanism, two highly-eccentric bodies
first undergo a severe orbital crossing and then form a short-period
planet via collision-merger. In the set of simulations performed to
date, this mechanism produces a short-period, terrestrial planet in
20$\%$ of runs.

As mentioned previously, the formation rate for short-period
terrestrial planets via a collision-merger process is only a
moderate 20\%. However, this low rate may be a result of the limits
imposed on our simulations by current computational capabilities,
which restrict our adopted population of embryos and planetesimals
to a few hundred objects with a total mass of only several times
that of the Earth. The resultant planetesimal disk in our
simulations is much smaller than that of the Minimum Mass Solar
Nebula ($\sim0.01$ $M_{\odot}$ within 30\,AU \citep{weid77,haya81})
-- which would also contain billions of small bodies. Increasing the
number of bodies and the total mass of the proto-planetary disk
would likely increase the efficiency with which this mechanism
produces short-period terrestrial planets.

In addition, it is worth noting that close-in planets emerge from
our simulations within a few million years. This is a significantly
shorter timescale than the billion years over which the Solar System
is thought to have undergone significant evolution. So, while
near-infrared observations of young cluster samples, indicate an
overall dust disk lifetime of $\sim6$\,Myr \citep{hais01}, the
planetary system will actually continue to evolve over much longer
timescales following the clearing of the dust and gas disk. During
this late stage of planetary formation, frequent orbital crossings
and huge impacts will occur, which are likely to significantly boost
the feasibility of collision-merger events producing short-period
terrestrial bodies.

The collision-merger scenario for the formation of short-period
planets  does not require perfect accretion. Rather it relies on the
collisions pushing the resultant body inward, so that the central
star can capture it as a short-period planet. In this sense, such a
mechanism could play a key role in throwing the largest fragments
resulting from severe impacts into short-period orbits. On the other
hand, given the diversity in the architectures of currently known
systems, exoplanets are likely to form through  a variety of
mechanisms rather than through a uniform dominant process (D. Lin
2009, private communication). Our simulation results show one
potential mechanism for the origin of short-period terrestrial
planets in a compact disk with two gas-giants, and may predict an
abundance of close-in bodies for this family.

\acknowledgments{We thank the anonymous referee for useful comments
and suggestions that helped to improve the contents. We also thank
J.E. Chambers, D.N.C. Lin and E. Kokubo for informative discussions
and insightful comments. J.H.J. is very grateful to the Australian
Academy of Sciences for the support of his stay at UNSW, and to
Chris Tinney and UNSW Astrophysics Department for their hospitality.
This work is financially supported by the National Natural Science
Foundation of China (Grants 10973044, 10833001, 10573040, 10673006,
10233020), the Natural Science Foundation of Jiangsu Province, and
the Foundation of Minor Planets of Purple Mountain Observatory.}




\end{document}